\documentclass[preprint,showpacs,english,amsmath,amssymb,superscriptaddress]{revtex4}
\usepackage[T1]{fontenc}
\usepackage[latin9]{inputenc}
\usepackage{amsmath}
\usepackage{graphicx}
\usepackage{color}
\usepackage{bm} %boldmath
\usepackage{babel}

\begin{document}

\title{Brownian motor in a granular medium}

\author{R. Balzan} \affiliation{Dipartimento di Fisica, Sapienza
  Universit\`{a} di Roma, Piazzale A. Moro 2, 00185 Roma, Italy}
\affiliation{Consiglio Nazionale delle Ricerche, Istituto dei Sistemi
  Complessi, sede di Tor Vergata, Via del Fosso del Cavaliere 100,
  00133 Roma, Italy}

\author{F. Dalton} \affiliation{Consiglio Nazionale delle Ricerche,
  Istituto dei Sistemi Complessi, sede di Tor Vergata, Via del Fosso
  del Cavaliere 100, 00133 Roma, Italy}

\author{V. Loreto} \affiliation{Dipartimento di Fisica, Sapienza
  Universit\`{a} di Roma, Piazzale A. Moro 2, 00185 Roma, Italy}
\affiliation{ISI Foundation, Viale Settimio Severo 65, Villa
  Gualino, I-10133 Torino, Italy}

\author{A. Petri}  \affiliation{Consiglio Nazionale delle Ricerche,
  Istituto dei Sistemi Complessi, sede di Tor Vergata, Via del Fosso
  del Cavaliere 100, 00133 Roma, Italy}

\author{G. Pontuale}  \affiliation{Consiglio Nazionale delle Ricerche,
  Istituto dei Sistemi Complessi, sede di Tor Vergata, Via del Fosso
  del Cavaliere 100, 00133 Roma, Italy}

\begin{abstract}
  In this work we experimentally study the behavior of a
  freely-rotating asymmetric probe immersed in a vibrated granular
  medium.  For a wide variety of vibration conditions the probe
  exhibits a steady rotation whose direction is constant with respect
  to the asymmetry. By changing the vibration amplitude and by
  filtering the noise in different frequency bands we show that the
  velocity of rotation does not depend only on the RMS acceleration
  $\Gamma$, but also on the amount of energy provided to two separate
  frequency bands which are revealed to be important for the dynamics
  of the granular medium: the first band governs the transfer of
  energy from the grains to the probe, and the second affects the
  dynamics by altering the viscosity of the vibro-fluidized material.
\end{abstract}
\maketitle

%\textbf{Introduction}

Exploiting spontaneous motion for producing energy has been a long
standing dream, the impossibility of which in equilibrium systems was
definitively established with the advent of thermodynamics showing
that \textit{perpetuum mobile} of the second kind is contrary to the
second principle for any closed system.  At the microscopic level this
conclusion relies on molecular chaos as discussed by Smoluchowski and
Feynman~\cite{reimann02,hanggi05}, though the laws of thermodynamics
do not prevent the possibility of extracting work from systems that
are isothermic and in a stationary state {\it but not at equilibrium}.
Such a state would require some sort of spontaneous symmetry breaking
and constitutes the subject of Brownian motors~\cite{reimann02}.
Experimental realizations, generally named thermal ratchets, have been
recently studied in different situations, ranging from the nano to the
micro scale~\cite{deanastumian02,gommers08}. Macroscopic realizations
have been developed in the field of granular media
(GM)~\cite{farkas99,vandermeer04}, a field of great relevance both in
applied science, for their industrial relevance, and in theory, for
their challenging properties and behavior~\cite{1-Review
  Granular}. Strictly speaking granular systems are athermal since the
motion of the elementary constituents of the medium, the grains, is
not affected by the ambient temperature.  It may therefore seem
contradictory to speak of thermal ratchets; however grains may gain
kinetic energy by external mechanical perturbations, and so be
ascribed a 'temperature' defined by their motion~\cite{3-Nature}, yet
never reach equilibrium because of the presence of inelastic
collisions and friction. Therefore, GMs can be used to implement
macroscopic realizations of thermal ratchets, both
experimentally~\cite{farkas99,vandermeer04,vandermeer07} and
theoretically \cite{baea04,costantini07,costantini08}.

The experimental realizations cited above were principally targeted towards
the observation of spontaneous collective oriented motion of the
grains, whereas in this work we wish to focus on the spontaneous
motion of an external asymmetric intruder {\em in the absence of any
  collective granular motion}.  Numerical simulations have
demonstrated that such a phenomenon is possible~\cite{costantini07}.
We present novel results showing that the chaotic granular motion can
indeed propel the asymmetric probe in a persistent direction.  More
specifically, the immersed asymmetric probe,
under suitable conditions of fluidization and
viscosity, exhibits consistent and significant motion in a
constant direction with respect to its asymmetry.  Under different
conditions the system can exhibit collective motion of the grains
(e.g. convection); however in these cases the symmetry breaking is
performed by the GM and so the direction of asymmetry of the probe is
irrelevant.

%\textbf{Experimental setup}

\begin{figure}
  \begin{centering}
    \includegraphics[width=7.2cm,keepaspectratio]{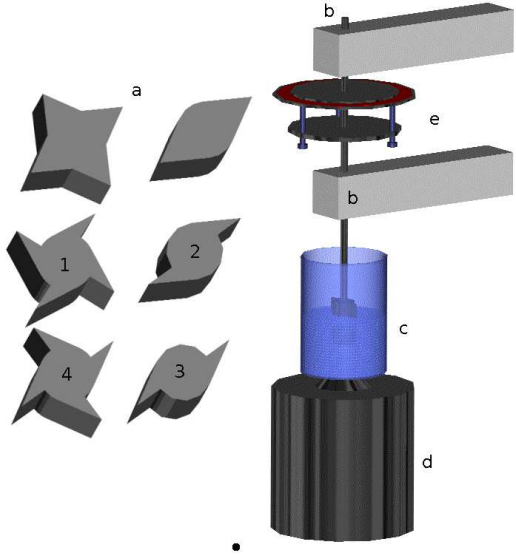}
    \par\end{centering}
  \caption{Example of the symmetric and asymmetric probes used
    (a). The probe is attached to a freely-rotating support (b) and is
    inserted into a beaker (c) containing the beads. The beaker is
    vibrated by a shaker (d) (Bruel \& Kjaer type 4809) and the 
    instantaneous angular position of the probe is registered by a
    rotary encoder (e) with resolution
    $1/500^\circ$.\label{fig:disegno}}
\end{figure}

The main components of the experimental apparatus are the granular
medium, the shaker and the probe as shown in Fig.~\ref{fig:disegno}.
The signal to the shaker is supplied by a function generator and is
amplified and filtered with an efficiency of $22-23\: db/oct$.  The
vibration was applied in the range from $10$ to $1000\: Hz$ and the
system acceleration ($0 < \Gamma < 4$, normalized to gravity $g$) is measured by an accelerometer
fixed to the vibrating stage of the shaker.  The probes, each 3~cm
tall, are immersed to a depth $H$ in the GM (glass beads of diameter
$2\: mm \pm 10\: \%$) which approximately half-fills a beaker, $90\:
mm$ diameter and $120\: mm$ tall.

%\textbf{General results}

It is well known that when a GM is shaken with a periodic signal, a
wide range of coherent motions can arise~\cite{aranson06}. Generally
speaking, such motions are due to a dynamic symmetry breaking in the
system~\cite{vandermeer04} and vary from the presence of isolated and
discontinuous collective movements of a minority of beads to major
effects like inclination of the surface or formation of a single
vortex in which the entire GM rotates.  In our system these coherent
motions typically arise when the shaker is excited by a pure square or
sine wave in a range of frequencies up to about $150-200\: Hz$.  For
example, well defined internal convective motion was observed which
caused the probe to rotate until it stops at a specific angle,
irrespective of the starting point.  Alternatively, in the presence of
vortices the probe was seen to rotate indefinitely in a constant
direction at roughly a constant velocity, with no dependence on either
the presence or the direction of the probe asymmetry - the probe
clearly follows the macroscopic collective motion of the granular
material.  However, above $200\: Hz$ or with white noise excitation,
only chaotic motion of beads was observed with no collective effect,
while over $1000\: Hz$ the granular material seems no longer to
respond to the agitation even for $\Gamma \gg 1$.  Hence white noise
was applied to the shaker in order to eliminate any convective or
collective motion of the GM, and this allows us to apply energy also
below $200\: Hz$ without triggering collective motion.  The use of
non-periodic signals requires a correct definition of
$\Gamma$~\cite{3-Nature}:
\begin{equation}
  \Gamma=\frac{\sqrt{2\left\langle\ddot{z}^{2}\left(t\right)\right\rangle}}{g}=\frac{\sqrt{2\left\langle\ddot{\zeta}^{2}\left(\nu\right)\right\rangle
    }}{g},
  % =\frac{\sqrt{2}RMS\left(\ddot{z}\left(t\right)\right)}{g}
  \label{eq:Gamma}
\end{equation}

\noindent in which $\ddot{z}\left(t\right)$ is the instantaneous
acceleration of the granular container and $\langle\dots\rangle$ is a
time average. $\Gamma$ is also expressed as a
function of the Power Spectral Density (PSD) where
$\ddot{\zeta}\left(\nu\right)$ is the Fourier transform of
$\ddot{z}(t)$. Equation~\eqref{eq:Gamma} reduces to the usual
definition of $\Gamma$ for sinusoidal excitations.  Other definitions
also exist which reduce to the usual $\Gamma$ for periodic signals,
however the above choice is motivated by the fact that
Eq.~\eqref{eq:Gamma} is proportional to the overall energy provided to
the granular medium, and furthermore that it yields the value
$\Gamma=1$ at an evident transition point in our experiments (this
transition is also observed at a calculated value of $\Gamma=1$ with
pure sinusoidal excitation).

Using white noise in a large enough frequency range eliminates
convective motions and the granular exhibits a uniform and homogeneous
behavior. Under these conditions it may be expected that no net
rotation of the probe would be observed due to the random nature of
the action of the granular material on the probe surface.  Contrarily,
however, there is an almost regular rotation of the probe when $\Gamma
> 1$, the direction of which depends on the orientation of the probe
asymmetry. No net rotation was observed using the symmetric
probes. The response of the probe at various excitation intensities
and frequency ranges has been studied with the observation that the
probe velocity is dependent on the upper and lower frequency limits of
the shaking signal.  The probe immersion depth $H$ into the GM has
also been varied.

%\textbf{Rotation velocity}

\begin{figure}
  \begin{centering}
    \includegraphics[width=7.2cm,keepaspectratio]{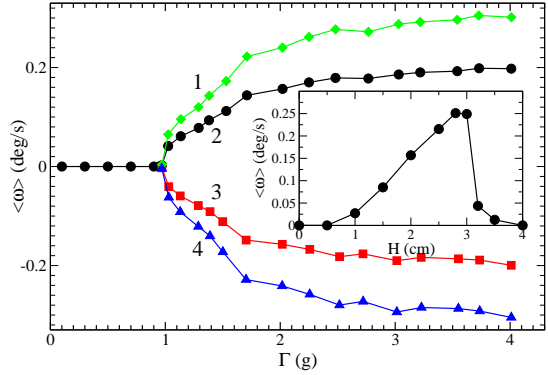}
    \par\end{centering}
  \caption{Probe average velocity as a function of the shaking acceleration
    for the $2$ and $4$ teeth asymmetric probes (numbered 1 to 4 as in
    fig.~\ref{fig:disegno}).  The vibration signal is white noise
    filtered in the range $15-300\: Hz$.  Inset: Average rotation velocity
    $\omega$ as a function of the immersion depth $H$ of the probe at
    $\Gamma=2$.  At $H=3\: cm$ the probe is fully immersed in the
    medium.}\label{fig:Velocity}
\end{figure}

Figure~\ref{fig:Velocity} presents a general result illustrating the
asymmetry-dependent mean rotation velocity $\langle\omega\rangle$ of
the probe as a function of the vibration intensity $\Gamma$.  Below
$\Gamma=1$ the velocity $\langle\omega\rangle=0$.  There is a clear
transition at $\Gamma=1$ at which point $\omega$ initially increases
rapidly before reaching an asymptotic value which is dependent on the
probe used (e.g.\ probes with 4 teeth rotate faster than those with
2), and the direction of its asymmetry - inverting the probe asymmetry
results in a simple inversion of $\omega$.  Notably, symmetric probes
show $\langle \omega\rangle =0\: \forall\: \Gamma$.

In the inset to Fig.~\ref{fig:Velocity} the average velocity of the probe for
various immersion depths $H$ of the probe into the granular material
is shown. For small $H<1\: cm$ there is no rotation and we assume that
the action of the GM on the probe is insufficient to overcome friction
in the support.

Increasing $H$ causes the probe surface to engage more completely with
the GM, and the velocity increases roughly linearly until $H\simeq
2.5\: cm$.  At this point the probe is almost fully immersed in the
medium and the velocity reaches a brief plateau.  Further immersion
$H>3\: cm$ causes the rotation velocity to fall quickly to $0$.  This
behavior suggests that the region of the GM which applies the
spontaneous torque to the probe is close to the surface.

%\textbf{Filtering bands}

By changing the intensity and the frequency range of the applied white
noise it may be seen that the behavior of the system is principally
determined by the energy provided in two major frequency bands.  We
consider a first series of experiments in which we apply a single
frequency band to the GM by applying a band-pass filter to white noise
from $f_1$ to $f_2\: Hz$.  Firstly by altering $15 < f_1 < 60\: Hz$
with fixed $f_2 = 300\: Hz$ then by fixing $f_1 = 15\: Hz$ and
altering $300 < f_2 < 900\: Hz$.  As the frequency band is widened,
the overall energy provided to the system would increase; to
investigate therefore only the change in frequency, the signal
intensity is reduced in order to maintain a constant overall energy
input, proportional to $\Gamma$ (the probe behaves similar to
Fig.~\ref{fig:Velocity} as a function of $\Gamma$).

\begin{figure}
  \begin{centering}
    \includegraphics[width=7.2cm,keepaspectratio]{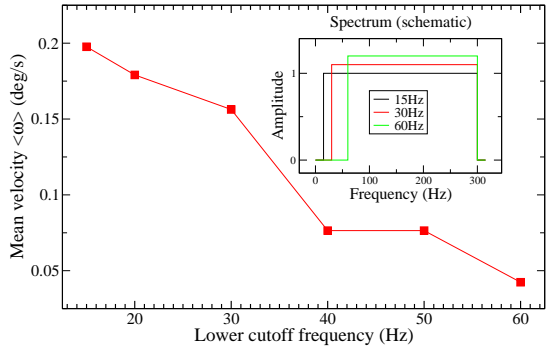}
  \par\end{centering}
\caption{\label{fig:low-freq-cutoff}The asymptotic value of the probe
  velocity (i.e.,\ $\langle \omega \rangle$ at $\Gamma=4$) while
  varying the lower cutoff $f_1$ of the band-pass filter from $15$ to
  $60$ Hz, necessitating an increase in the filter amplitude in order
  to maintain a constant $\Gamma$.  The mean velocity decreases
  approximately proportional to $f_1$ with a step between $30$ and
  $40$ Hz, coincident with the absorption peak in the response
  function.}
\end{figure}

For different $f_1$ and fixed $\Gamma$ the probe approaches a
different asymptotic value.  Figure~\ref{fig:low-freq-cutoff} displays
this asymptotic value of the mean probe angular velocity $\langle
\omega \rangle$ (taken at $\Gamma=4$) as a function of $f_1$, the
lower band-pass cutoff.  It is clear that the $\langle \omega \rangle$
decreases slowly from $f_1 = 15\: Hz$ to $f_1=30\: Hz$, from $30$ to
$40\: Hz$ there is a jump which corresponds to an absorption peak in
the apparatus response function~\cite{BALZAN-THESIS}, due to the
granular material efficiently absorbing energy from the vibration and
so mobilizing the probe.  For $f_1>40\: Hz$ $\langle \omega \rangle$
continues to decrease slowly.  Importantly, this jump between $30$ and
$40\: Hz$ is independent of the value of the higher cut-off frequency
$f_2$ (in this example fixed at $300\: Hz$, though fixed at up to
$900\: Hz$ is other experiments).  Thus we conclude that the frequency
band from roughly $30$ to $40\: Hz$ is crucial to generating motion of
the probe.

\begin{figure}
  \begin{centering}
    \includegraphics[width=7.2cm,keepaspectratio]{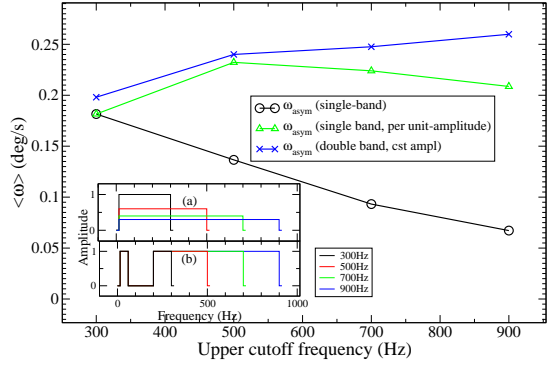}
    \par
  \end{centering}
  \caption{\label{fig:high-freq-cutoff}The value of the probe velocity
    (measured at $\Gamma=4$) while varying the upper cutoff $f_2$ of
    the band pass filter. This was performed both with a single band,
    maintaining the overall signal energy by decreasing the amplitude
    as appropriate, and with a double-band at constant amplitude (see inset).  In
    the first case, the velocity decreases (black circles) displaying a linear
    behavior with respect to the amplitude (green triangles).  In the second
    case, increasing 7-fold the energy in the upper band produces only
    a slight increase in the mean probe velocity (blue $\bm{\times}$); this is in
    contrast to the result in Fig.~\ref{fig:low-freq-cutoff} where a
    reduction of approx.\ 15\% resulted in a 75\% reduction in the
    velocity.}
\end{figure}

We now fix $f_1=15\: Hz$ and increase the upper band-pass cutoff
frequency $300 < f_2 < 900\: Hz$, still maintaining a constant overall
energy input by reducing the signal amplitude as necessary, as
schematized in the inset to Fig.~\ref{fig:high-freq-cutoff}(a).  The
main Fig.~\ref{fig:high-freq-cutoff} demonstrates that the mean
velocity actually {\em decreases} as the frequency band is widened
(black circles).  This may seem contrary to Fig.~\ref{fig:Velocity} but
in fact is a direct consequence of Eq.~(\ref{eq:Gamma}) and the reduction
in signal intensity, specifically in the $30 - 40\: Hz$ band, which is
required to maintain a constant $\Gamma$.  In fact, if one divides the
velocity obtained by the amplitude applied (relative to that with
$f_2=300\: Hz$), one obtains an almost constant velocity (green
triangles), indicating linearity of $\omega$ with $\gamma$.

To overcome this effect, we consider a final series of experiments in
which the amplitude of the signal sent to the shaker is kept constant.
In order to supply sufficient energy in the bands which interested us
($15-60\: Hz$ and $200-900\: Hz$), it was necessary to apply a
band-reject from $60$ to $200\: Hz$; the spectrum applied to the
shaker is schematized in the inset to
Fig.~\ref{fig:high-freq-cutoff}(b).  In this manner a constant energy
input in the lower band is maintained while the upper band can be
independently widened.

In Fig.~\ref{fig:high-freq-cutoff} we show the variation of the
asymptotic probe velocity as the upper band is widened from $300$ to
$900\: Hz$ (blue $\bm{\times}$).  The velocity is seen to increase only
slightly despite the large increase in the band width.

When taken together with the observation that motion ceases when
energy is removed from the lower-band {\em irrespective of the upper
  band}, then the logical conclusion is that the upper frequency band
is not responsible for the spontaneous torque causing the rotation.
The slight increase in velocity, therefore, may be ascribed to a
reduction in the viscosity of the GM; indeed we imagine a
high-frequency vibration merely weakening the contacts between grains
which, however, remain caged in place by their neighbors, whereas a
low-frequency motion actually imparts macroscopic motion to the
grains.  Such a phenomenon has already been observed in~\cite{3-Nature}.

The viscous torque $M_g$ can be written as the product of the probe
viscous coefficient in the GM $\nu_B$ and the probe velocity $\omega$:
\begin{equation}
  M_{g}=-\nu_{B}\omega.\label{eq:resistenza}
\end{equation}
where $\nu_B$ is explicitly dependent on the excitation band, and in fact
is inversely proportional to $\Gamma$~\cite{3-Nature}:
\begin{equation}
  \nu_{B} = \frac{a(PSD(f))}{\Gamma}.\label{eq:nu di Gamma}
\end{equation}

\noindent where $a(PSD(f))$ is some function of the spectrum applied
to the shaker.

Thus we infer that, when the lower cut-off is increased, $\Gamma$ is
drastically reduced, which both increases the viscosity, and reduces
the spontaneous torque $F$ generated by the GM.  When the upper
cut-off is increased, however, $\Gamma$ remains almost constant but a
decrease in viscosity is observed due to the presence of the term
$a(PSD(f))$ in Eq.~(\ref{eq:nu di Gamma}).

%\textbf{Discussion}

In summary, the origin of the observed behavior can be traced back to
both the mechanical absorption of the GM and its fluidization.  In the
case of $\Gamma<1$ rotation does not occur as the GM is insufficiently
mobilized and maintains a high viscous coefficient $\nu_B$.  The
reduction of $\nu_{B}$ with $\Gamma>1$ confers sufficient mobility to
the probe and so rotation occurs.  It is clear from
Fig.~\ref{fig:high-freq-cutoff} that a slight decrease of $\nu_{B}$,
due to the widening band produces a slight increase in the probe
velocity (blue curve).

During experimentation, the granular medium is clearly seen to gain
considerable freedom near the top surface and the beads closest to the
faces of the probe tend to move away from the probe, reaching zones of
the granular surface which hinder less the probe rotation. In doing so
the beads leave the probe radially and, because of the asymmetry,
contribute some net momentum to the probe and trigger rotation.  The
gained momentum is smaller for the faces parallel to the radii of the
probe so the direction of rotation is that in which these faces move
ahead.  This is consistent with the observation that a completely
immersed probe, no longer in contact with the top surface, ceases to
rotate Fig.~\ref{fig:Velocity} (inset).  In this respect the results
are different from those reported in~\cite{angelani09} for the
simulation of a Brownian motor driven by bacteria.

Given the results obtained above, the following equation of motion for
the probe may be written:
\begin{equation}
  I\dot{\omega}=F-R-\nu_{B}\omega+N
  \label{eq:eq moto}
\end{equation}
where $I$ is the moment of inertia of the rotating elements (the
probe, its support and any dragged GM~\cite{note}), $F$ is the net
torque exerted by the GM on the probe, $R$ is the frictional torque of
the support, considered constant, $\nu_{B}\omega$ represents the
viscous resistance torque due to the granular~\cite{2-Tesi
  Mayor,3-Nature}, and $N$ is a noise term indicating fluctuations of
$F$ and is assumed to have zero time average, $\left\langle N(t)
\right\rangle=0$.

%\textbf{Conclusions and outlook}

The main achievement of the experiment reported here has been the
experimental observation of the Brownian motor effect at a
macroscopic, $cm$, scale.  The observations are in general agreement
with other works related to hydrodynamic viscosity~\cite{2-Tesi
  Mayor,3-Nature} and computer simulations~\cite{Simulation}.  
A characteristic mechanical resonance is found in the range $30-40\:
Hz$, similar to that found by numerical experiments~\cite{Simulation}
which, when excited, triggers spontaneous rotation of the immersed
asymmetric probe.  Excitation at higher frequency reduces the
viscosity of the GM, but does not in itself trigger rotation.

{{\em Acknowledgments}} V.L. wishes to thank Patrick Mayor and
Gianfranco D'Anna with whom a preliminary version of the experimental
setup was conceived.  R.B. thanks Simone Franchini and Valerio Paladino
for useful discussions and suggestions.  F.D.\ gratefully acknowledges
financial support from E.U.\ Nest/Pathfinder project TRIGS under contract
no.\ NEST-2005-PATH-COM-043386.


\begin{thebibliography}{99}

\bibitem{reimann02} P. Reimann, Phys. Rep. \textbf{361}, 57 (2002);

\bibitem{hanggi05} P. H\"{a}nggi, F. Marchesoni and Franco Nori,
  Ann. Phys.  \textbf{14}, 51 (2005);

\bibitem{deanastumian02} R. Dean Astumian and P. H\"{a}nggi,
  Phys. Today \textbf{55}, 32 (2002);

\bibitem{gommers08} R. Gommers, V. Lebedev, M. Brown, and F. Renzoni,
  Phys. Rev. Lett. \textbf{100}, 040603 (2008);

\bibitem{farkas99} Z. Farkas, P. Tegzes, A. Vukics, and T. Vicsek,
  Phys. Rev. \textbf{E 60},7022 (1999);

\bibitem{vandermeer04} D. van der Meer, P. Reimann, K. van der Weele
  and D. Lohse, Phys. Rev. Lett. \textbf{92}, 184301 (2004);

\bibitem{1-Review Granular} H. M. Jaeger and S. R. Nagel,
  Rev. Mod. Phys. \textbf{68}, 1259 (1996);

\bibitem{3-Nature} G. D'Anna, P. Mayor, A.Barrat, V. Loreto and
  F. Nori, Nature 4\textbf{24}, 909 (2003);


  % \bibitem{vandermeer06} Devaraj van der Meer, Ko van der Weele, and
  %   Peter Reimann, Phys. Rev.\textbf{E 73}, 061304 (2006).

\bibitem{vandermeer07} D. van der Meer, K. van der Weele, P. Reimann
  and D. Lohse, J. Stat. Mech P07021 (2007);

\bibitem{note} The moments of inertia of a solid dragging a granular
  medium is increased by the motion of the dragged grains themselves
  as reported e.g. in~\cite{francois02,baldassarri06}

\bibitem{francois02} B. Francois, F. Lacombe, and H. J. Herrmann,
  Phys. Rev. \textbf{E 65}, 031311 (2002);

\bibitem{baldassarri06} A. Baldassarri, F. Dalton, A. Petri et al.,
  Phys. Rev. Lett. \textbf{96}, 118002 (2006);

\bibitem{baea04} A. J. Bae, W. A. M. Morgado J. J. P. Veerman and
  G. L. Vasconcelos, Physica \textbf{A 342}, 22 (2004);

\bibitem{costantini07} G. Costantini, U. Marini Bettolo Marconi and
  A. Puglisi, Phys. Rev. E 75 061124 (2007);

\bibitem{costantini08} G. Costantini, U. Marini Bettolo Marconi,
  A. Puglisi, Europhys. Lett. \textbf{82}, 50008 (2008);

  % \bibitem{chepelianskii08} A. D. Chepelianskii, M. V. Entin,
  %   L. I. Magarill, and D. L. Shepelyansky Physical Review E 78
  %   041127 (2008)

\bibitem{aranson06} I.S. Aranson and L. S. Tsimring,
  Rev. Mod. Phys. \textbf{78}, 641 (2006);

\bibitem{2-Tesi Mayor} P. Mayor, Fluid and glassy phases of vibrated
  granular matter studied with a torsion oscillator, PhD Thesis N3334
  (2005) EPFL (CH);


\bibitem{Simulation} J. A. C. Gallas, H. J. Herrmann and
  S. Sokolowski, Phys. Rev. Lett. \textbf{69}, 1371 (1992);

\bibitem{angelani09} L. Angelani, R. Di Leonardo and G. Ruocco,
  Phys. Rev. Lett. \textbf{102}, 048104 (2009).

\bibitem{BALZAN-THESIS} R. Balzan, Studio sperimentale di mezzi
  granulari: motore browniano, Master Thesis (2008), University of Rome "La Sapienza",
  Italy.

\end{thebibliography}
\end{document}